\documentclass[doublecol]{epl2}

\title{Anti-correlation and subsector structure in financial systems}

\author{X.F. Jiang and B. Zheng\footnote{corresponding author; email: zheng@zimp.zju.edu.cn}}
\shortauthor{X.F. Jiang and B. Zheng}

\institute{
Department of Physics, Zhejiang University, Hangzhou 310027,
P.R. China}

\pacs{89.65.Gh}{Economics; econophysics, financial markets, business
and management } \pacs{89.75.-k}{Complex systems}

\abstract{ With the random matrix theory, we study the spatial
structure of the Chinese stock market, American stock market and
global market indices. After taking into account the signs of the
components in the eigenvectors of the cross-correlation matrix, we
detect the subsector structure of the financial systems.
The positive and negative subsectors are anti-correlated each other
in the corresponding eigenmode. The subsector structure is strong
in the Chinese stock market, while somewhat weaker in the American
stock market and global market indices. Characteristics of the subsector structures
in different markets are revealed.}

\begin{document}

\maketitle

\section{Introduction}

In recent years, much attention of physicists has been attracted to
the financial dynamics, which exhibits various collective behaviors
\cite{man95,gop99,gia01,bou01,qiu06,she09,she09a,pod09,pod10}.
Statistical properties of price fluctuations and cross-correlations
between individual stocks are of great
interest, not only for quantitatively unveiling the complex
structure of the financial systems, but also practically for the
asset allocation and portfolio risk estimation
\cite{far99,man00,bou03}. The probability distribution of price
returns usually exhibits a power-law tail, and represents the robust
characteristics in stock markets \cite{lux96,ple99a,pan07}, while
the higher-order time correlations and interactions between stocks
are less universal \cite{qiu06,pan07a,she09,she09a}.
In some cases, price returns may also show a Poisson-like distribution \cite{bap00,bap02}.

It is an important and challenging topic to explore the 'spatial'
structure in financial systems. For example, the hierarchical
structure of stock markets has been investigated through the minimal
spanning tree method and its variants
\cite{man99,bon03,tum05,tum07,ast10}. With the random matrix theory
(RMT), business sectors and topology communities may be identified
\cite{ple02a,pan07a,uts04,gar08}. The RMT method was firstly
developed in the complex quantum systems where the interactions
between subunits are unknown \cite{meh91,guh98}. The structure of business sectors have been
examined for mature markets such as the New York Stock Exchange
(NYSE) and the Korean Stock Exchanges
\cite{lal99,ple99,ple02a,uts04,qiu10,oh11}, and also for some
emerging markets such as the National Stock Exchange in India
\cite{pan07a}. Very recently, the RMT method was applied to identify
the dominant eigenmodes in the indices of the industrial production
\cite{iye11}. In particular, one has investigated the structure of
interactions between stocks for the Chinese stock market based on
the RMT method \cite{she09}. As an important emerging market, the
Chinese market exhibits stronger cross-correlations than
the mature ones. At the same time, the effect of the
standard business sectors is weak in the Chinese market. Instead,
unusual sectors such as ST and Blue-chip sectors are
detected.

In this paper, with the RMT method, we aim at further understanding
of the spatial structure. Our
observation is that the components in an eigenvector of the
cross-correlation matrix may show positive
and negative signs. To the best of our knowledge, what roles the
signs of the components play has not been explored. Our main finding
is that the signs of the components in an eigenvector may classify a
sector into two subsectors, which are anti-correlated each other
within this eigenmode. This goes beyond what one may gain
with standard methods such as the minimal spanning tree and
its variants in the analysis of the 'spatial' structure in financial systems \cite{man99,bon03,tum05,tum07,ast10}.

\section{Methods and basics}

We have collected the daily data of $259$ stocks traded in the
Shanghai Stock Exchange (SSE) from Jan., 1997 to Nov., 2007, in
total, 2633 days. The daily data of 259 stocks in the NYSE are from
Jan., 1990 to Dec., 2006, in total, 4286 days. Meanwhile, we have
collected the daily data of a set of $66$ financial indices,
including $57$ indices in stock markets and $9$ treasury bond rates
in US from Sep., 1997 to Oct., 2008, in total, $2669$ days. We name
the 66 indices the {\it global market indices} (GMI). The data of
the SSE are taken from 'Wind Financial Database'
(http://www.wind.com.cn) and the data of the NYSE and GMI are from
'Yahoo Finance' (http://finance.yahoo.com).

We assume that the price is the same as the preceding day
\cite{wil07}, if the price of a stock is absence in a particular
day. It has been pointed out that the missing data do not result in
artifacts \cite{pan07a}. All markets
concerned in this paper have normal trading sessions in all days of
the week, except for Saturdays, Sundays and holidays declared in
advance, excluding the Egyptian and Tel Aviv Stock Exchange. For the
latter two stock markets, trading takes place from Sunday to
Thursday. For the alignment of the time series, we simply move the
data on Sunday to Friday for these two markets.
For comprehensive understanding of the cross correlation of
financial markets, the bond rates in US are also added in our
analysis, which include $9$ indices ranging from the $3$ month to
$20$ year rates.

We define the logarithmic price return of the \textit{$i$}-th stock
over a time interval $\Delta t$ as
\begin{equation}
R_{i}\left(t\right)\equiv ln\, P_{i}\left(t+\Delta
t\right)-ln\, P_{i}\left(t\right),\label{eq:1}\end{equation} where
$P_{i}\left(t\right)$ represents the price of the stock price at
time $t$. To ensure different stocks with an equal weight, we
introduce the normalized price return
\begin{equation}
r_i\left(t\right)=\frac{R_{i}(t)-\langle R_{i}(t)\rangle}{\sigma_{i}},\label{eq:2}\end{equation}
where $\langle \cdots\rangle$ is the average over time $t$, and
$\sigma_{i}=\sqrt{\langle R_{i}^{2}\rangle-\langle R_{i}\rangle^{2}}$ denotes the standard
deviation of $R_{i}$. Then, the elements of the
cross-correlation matrix $C$ are defined by the equal-time correlations
\begin{equation}
C_{ij}\equiv\langle r_{i}(t)r_{j}(t)\rangle.\label{eq:3}
\end{equation}
By the definition, $C$ is a real symmetric matrix with $C_{ii}=1$,
and $C_{ij}$ is valued in the domain $\left[-1,1\right]$.

The mean value $\overline C_{ij}$ of the elements for the SSE is $0.37$, much
large than $0.16$ and $0.26$ for the NYSE and GMI respectively.
It confirms that stock prices in emerging markets are more
correlated than mature ones \cite{mor00,pod06,pan07a}. The correlation between financial indices in the GMI
is smaller than that of the SSE, but bigger than that of the NYSE.

We now compute the eigenvalues of the cross-correlation matrix
$C$, in comparison with those of the so-called \textit{Wishart}
matrix, which is derived from non-correlated time series. Assuming
$N$ time series with length $T$, and in the large-$N$ and large-$T$
limit with $Q\equiv T/N\,\geq 1$, the probability distribution
$P_{rm}(\lambda)$ of the eigenvalue $\lambda$ for the Wishart matrix
is give by \cite{dys71,sen99}
\begin{equation}
P_{rm}(\lambda)=\frac{Q}{2\pi}\frac{\sqrt{\left(\lambda_{max}^{ran}-\lambda\right)\left(\lambda-\lambda_{min}^{ran}\right)}}{\lambda},
\label{eq:4}\end{equation} with the upper and lower bounds
\begin{equation}
\lambda_{min\left(max\right)}^{ran}=\left[1\pm\left(1/\sqrt{Q}\right)\right]^{2}.
\label{eq:5}\end{equation}
For a dynamic system, large eigenvalues of the cross-correlation matrix,
which deviate from $P_{rm}\left(\lambda\right)$, imply that there
exists non-random interactions. In fact, in both mature and emerging
stock markets, the bulk of the eigenvalue spectrum $P(\lambda)$ of
the cross-correlation matrix is similar to $P_{rm}(\lambda)$ of the
Wishart matrix, but some large eigenvalues deviate significantly
from the upper bound $\lambda_{max}^{ran}$. This scenario looks
similar for the GMI. Let us arrange the large eigenvalues in the
order of $\lambda_{\alpha}>\lambda_{\alpha+1}$. As shown in
table~\ref{table-basic-stat}, the largest eigenvalue $\lambda_{0}$
of the SSE (China) is $97.33$, about $56$ times as large as the
upper bound $\lambda_{max}^{ran}$ of $P_{rm}(\lambda)$, while
$\lambda_{0}$ of the NYSE (US) and GMI is $45.61$ and $21.53$, about
$29$ and $16$ times as large as $\lambda_{max}^{ran}$ respectively.

According to the previous works \cite{gop01,ple02a,pan07a,she09},
the large eigenvalues deviating from the bulk
correspond to different modes of motion in stock markets. The
components in the eigenvector of the largest eigenvalue
$\lambda_{0}$ are uniformly distributed. Therefore, the largest
eigenvalue represents the market mode, which is driven by
interactions common for stocks in the entire market. The components
in the eigenvectors of other large eigenvalues are localized. A
particular eigenvector is dominated by a sector of stocks, usually
associated to a business sector.
By $u_i\left(\lambda_{\alpha}\right)$, we denote the component of the $i$-th stock in the eigenvector of $\lambda_\alpha$.
To identify the sector, one may
introduce a threshold $u_{c}$, to select the dominating components
in the eigenvector by $\mid
u_i\left(\lambda_{\alpha}\right)\mid\geq u_{c}$ \cite{she09}.
The threshold $u_{c}$ is determined by two criteria. Firstly, if
the matrix is random, $<|u(\lambda)|> \sim 1/\sqrt{N}$
for every eigenmode. Therefore, $u_{c}$ should be larger than $1/\sqrt{N}$.
Secondly, $u_{c}$ should not be too large, otherwise there would be not
so many stocks in each sector.

In this paper, we show that the components in an
eigenvector may carry positive and negative
signs, and the components with opposite signs are anti-correlated within
this eigenmode. Inspired by this observation, we investigate the
subsector structure of the financial markets, by taking into account
the signs of the components. In other words, we separate a sector
into two subsectors by two thresholds $u^{\pm}_c=\pm u_c$:
$u_i\left(\lambda_{\alpha}\right)\geq u^+_{c}$ and
$u_i\left(\lambda_{\alpha}\right)\leq u^-_{c}$, which correspond to
the positive and negative subsectors respectively.

\section{Subsectors}

According to reference~\cite{she09}, standard
business sectors can
hardly be detected in the SSE (China). Instead, one finds that there exists three
unusual sectors, i.e., the ST, Blue-chip and SHRE sectors,
corresponding to the second, third and forth largest eigenvalues
respectively. What are the dominating stocks for the eigenvectors of
other large eigenvalues remains puzzling.

In the SSE, a company will be specially treated if its financial
situation is abnormal. Then a prefix of the acronym ``ST'' will
be added to the stock ticker.
The acronym ``ST''  will be removed
when the financial situations becomes normal. In
reference~\cite{she09}, the so-called ST sector consists of the ``ST''
stocks. On the other hand, the Blue-chip sector is
referred to those companies with a national reputation, and with good performance, i.e., a
reasonable positive profit in a period of time. Meanwhile, the SHRE
represents the companies registered in Shanghai with the real estate
business.

Now we introduce two thresholds $u^{\pm}_c=\pm u_c$ to separate the
dominating components in an eigenvector into two parts, i.e.,
$u_i\left(\lambda_{\alpha}\right)\geq u^+_{c}$ and
$u_i\left(\lambda_{\alpha}\right)\leq u^-_{c}$, which are referred to
the {\it positive} and {\it negative subsectors} respectively. With this method,
we are able to identify the subsectors of the
SSE up to the seventh largest eigenvalue $\lambda_{6}$, and to
achieve deeper understanding on the unusual sectors such as the ST
and Blue-chip sectors. The results are shown in
table~\ref{table-sector-SSE}. The market mode described by the
largest eigenvalue $\lambda_{0}$ is not included in the table, where
all components in the eigenvector posses a same sign.

The negative components in the eigenvector of the second largest
eigenvalues $\lambda_{1}$ are dominated by the ST stocks. With the
threshold $u^-_{c}=-0.10$, for example, $23$ dominating stocks are
selected, and $20$ of them are the ST stocks. Therefore this
subsector is called the ST subsector. For the positive components in
the eigenvector of $\lambda_{1}$, we could not identify a common
feature for the dominating stocks. In fact, as the threshold
$u^+_{c}$ increases, the number of the dominating stocks shrinks.
For example, with the threshold $u^-_{c}=-0.10$, there are only $7$
dominating stocks, and half are also the ST stocks. In
reference~\cite{she09}, therefore, the whole sector of $\lambda_{1}$
is called the ST sector. The negative components in the eigenvector
of the third largest eigenvalue $\lambda_{2}$ well define the high
technology subsector, while the positive ones are dominated by the
traditional industry stocks. Stocks in both subsectors are the
Blue-chip stocks. Therefore, these two subsectors together are
ascribed to the Blues-chip sector in reference~\cite{she09}. For the
fourth largest eigenvalue $\lambda_{3}$, the SHRE sector detected in
reference~\cite{she09} splits into two subsectors, i.e., the SHRE
and ST subsectors. Consistent with the result in
reference~\cite{she09}, half of the ST stocks are also the
SHRE stocks. But the ST stocks here are different from those for
$\lambda_{1}$.

In reference~\cite{she09}, the sector structure is explored only up
to $\lambda_{3}$. With the exploration of the subsector structure, we are
able to step further. For
$\lambda_{4}$, the positive and negative subsectors are identified
to be the weakly and strongly cyclical industry respectively. The
former includes the stocks which fluctuate little with the economic
cycle, such as the daily consumer goods and services, while the
latter is blooming or depressing with the economic cycle, including
the basic materials and energy resources. The positive components in
the eigenvectors of $\lambda_{5}$ and $\lambda_{6}$ are dominated by
the finance and non-daily consumer subsectors, although the negative
ones remain unknown.

Taking into account the signs of the components in the eigenvector,
one may explore the subsector structure in the SSE up to
$\lambda_{6}$. A number of standard business subsectors such as the
high technology and finance are also observed. But the SSE is indeed dominated by unusual sectors and subsectors such
as the ST, Blue-chip, traditional industry, SHRE, weakly and strong
cyclical industry.
In China, the companies are not operated strictly within the registered
business. Therefore, standard business subsectors are rarely observed.
From the view of the behavioral psychology, the investors
in China are extraordinarily looking at the performance of the
companies and the dominating business and areas, etc.
Therefore, unusual sectors such as the ST, Blue-chip, and SHRE emerge.

For comparison, we also apply this method to study the
subsector structure in the NYSE (US). The results are listed in
table~\ref{table-sector-NYSE}. The subsector structure in
the NYSE is somewhat different from that in the SSE. From general
believing, the standard business subsectors should dominate the
eigenvectors of the large eigenvalues. Additionally, it would be
expected that there exists only one dominating subsector in an
eigenvector, probably under certain conditions, e.g., when the total
number of stocks is sufficiently large. To clarify these issues, our
results are presented up to the thresholds $u^\pm_{c}=\pm0.12$. As
shown in table~\ref{table-sector-NYSE}, most subsectors are indeed
the standard business subsectors. For
$\lambda_{1}$, $\lambda_{2}$, $\lambda_{6}$ and $\lambda_{11}$, only
one dominating subsector remains for sufficiently large thresholds
$u^\pm_{c}$. For $\lambda_{3}$, $\lambda_{7}$, $\lambda_{8}$ and
$\lambda_{9}$, however, there are two dominating subsectors. For our
dataset of the NYSE, our method does also provide a deeper
understanding on the spatial structure.

Finally, as shown in table~\ref{table-sector-GMI}, the subsectors in the
GMI can be identified with the threshold $u_{c}=\pm0.15$,
exclusively in terms of the {\it areas} to which the indices belong.
Different from the SSE and NYSE, the eigenvector of the largest
eigenvalue $\lambda_{0}$ of the GMI does not describe the so-called
'market mode', which represents the global motion of the financial
system. This may reflect the fact that all
the financial markets in the world have not been in such a unified
status. The first, second and third largest eigenvalues correspond
to the US, Asia-Pacific and Bond sectors, with only a single dominating
subsector. The US sector mainly consists of the indices in
US, except for the GSPTSE from Canada and GDAXI from Germany.
This result reflects that US is
the dominating economy in the world. From $\lambda_{3}$ to
$\lambda_{7}$, there emerge two dominating subsectors. One important feature of the
subsector structure is that the indices in the mainland of China or
in Hongkong always form an independent subsector. On the other hand,
the US bond rates do not mix with the indices in stock markets. For
$\lambda_{6}$, the short-term bond rates and long-term bond rates
are separated into the positive and negative subsectors
respectively.

\section{Anti-correlation between subsectors}

{\it What is the physical meaning of the positive and negative subsectors?} The
cross-correlation between two stocks can be written as
\begin{equation}
C_{ij}=\sum_{\alpha=1}^{N}\lambda_{\alpha}C_{ij}^{\alpha},\quad
C_{ij}^{\alpha}=u_{i}^{\alpha}u_{j}^{\alpha} \label{eq:7}
\end{equation}
where $\lambda_{\alpha}$ is the $\alpha$-th eigenvalue,
$u_{i}^{\alpha}$ is the $i$-th component in the eigenvector of
$\lambda_{\alpha}$, and $C_{ij}^{\alpha}$ represents the
cross-correlation in the $\alpha$-th eigenmode. In other words, the
cross-correlation between two stocks can be decomposed into those
from different eigenmodes. Since the eigenvalue $\lambda_{\alpha}$
is always positive, it gives the weight of the $\alpha$-th
eigenmode, and the sign of $C_{ij}^{\alpha}$ is essential in the
sum. According to Eq.~(\ref{eq:7}), $C_{ij}^{\alpha}$ is
positive if the components $u_{i}^{\alpha}$ and $u_{j}^{\alpha}$
have the same sign in a particular eigenmode. Otherwise, it is
negative. When $C_{ij}^{\alpha}$ is negative, two stocks are
referred to be {\it anti-correlated in this eigenmode}: when the
price return of the $i$-th stock is positive, the price return of
the $j$-th stock tends to be negative in the statistical sense.
Therefore, all stocks in a same subsector are positively correlated
in this eigenmode, while the stocks in different subsectors are
anti-correlated. This is the physical meaning of the
subsectors. For the NYSE, however, only a number of sectors split into two
subsectors. This suggests that the spatial structure and interactions among the stocks in
the SSE are more complicated.

Let us examine
some examples in the SSE. The sector of $\lambda_{2}$ is
composed of the traditional industry and high technology subsectors.
The former represents those traditional industry companies with a long-term
and stable interest, but a lower asset risk and expected revenue, while the latter includes the high technology companies
with novel business and conceptions, but a higher asset risk and
potential profit.
In a particular period, for example, the stock market is uncertain, and investors prefer the
traditional industries with a lower risk, then their stock prices rise up higher than those of the high technology companies.
In another period, however, the stock market is booming, and the situation is reverse.
Thus, these two subsectors are anti-correlated in
the eigenmode of $\lambda_{2}$. The sector of $\lambda_{4}$ consists
of the weakly and strongly cyclical industry subsectors.
Both subsectors are unusual, but their anti-correlation seems obvious.
The weakly and strongly cyclical industries are weakly and strongly correlated with the macro-economy environment respectively.
Thus, investors prefer the strongly cyclical industry when the macro-economy is booming. Instead, investors rather choose the weakly cyclical industry when the macro-economy declines.
In reference~\cite{she09}, the sector of $\lambda_{3}$ is identified
as the SHRE sector. Now this sector splits into the ST and SHRE
subsectors. In fact, half of the ST
stocks also belongs to the SHRE stocks. This suggests that the
investors care much the normal and abnormal financial situation,
even for the SHRE companies.

In the NYSE, the subsector structure of $\lambda_{3}$, $\lambda_{7}$
and $\lambda_{9}$ is understandable. The daily consumer goods and
services are considered as the traditional industries, while the
high technology and finance belong to another category. These two
sorts of stocks may show an anti-correlation, consistent with the
subsector structure of $\lambda_{2}$ in the SSE. For $\lambda_{2}$
with the threshold $u^{\pm}_c=0.08$, a weak subsector structure is
observed in the NYSE. From their intrinsic properties, the
daily consumer goods and basic materials are classified as the
weakly and strongly cyclical industries respectively. This is
similar to the case of $\lambda_{4}$ in the SSE. For $\lambda_{8}$,
the subsectors may be also explained along the lines above.

In the GMI, two examples are typical. The first one is the
subsectors of $\lambda_{5}$, where all components except for one
are the indices in the American stock markets. one subsector is
composed of the IIX, IXIC, NDX, NWX, PSE and SOXX.  Most of these
indices are related to the information technology, semiconductor
industry, internet industry, etc, with a potentially high payoff and
asset risk. The other subsector consists of the XMI, DJA, DJI, DJU
and DJX. Most of them are for the weighted and traditional
companies, which share the general feature of a stable currency flow
and mature business mode, but a lower profit. These two subsectors
are anti-correlated in this eigenmode. The second example is the
subsectors of $\lambda_{6}$, which are obviously anti-correlated for
they are just short-term and long-term bond rates in US. For
$\lambda_{3}$, $\lambda_{4}$ and $\lambda_{7}$, the subsector
structure indicates that the stock markets in China are somewhat
special.

To quantitatively measure the
anti-correlation between the positive and negative subsectors,
we construct the combinations of stocks in the two subsectors, $I^{\pm}_\alpha(t)=\sum_i u^{\pm}_{i}(\alpha)r_{i}(t)$, and compute
the cross-correlation
\begin{equation}
C_{+-}(\alpha)=\langle I^{+}_\alpha(t)I^{-}_\alpha(t)\rangle. \label{eq:8}
\end{equation}
Here $u^{\pm}_{i}(\alpha)$ is the $i$-th positive or negative
component in the $\alpha$-th eigenmode selected by the threshold, e.g., $u^{\pm}_c =\pm 0.08$. In
figure~\ref{fig.1}, the cross-correlation $C_{+-}(\alpha)$ is shown for the SSE and NYSE,
in comparison with that between two random combinations of stocks.
This result is not qualitatively sensitive to whether one introduces the thresholds
$u^{\pm}_c$ to select the dominating components.

In figure~\ref{fig.1}, we observe that $C_{+-}(\alpha)$
monotonically increases, and gradually approaches that for two
random combinations of stocks.
$C_{+-}(\alpha)$ computed with $I^{\pm}_\alpha(t)$ is smaller than that with two random combinations of stocks
because of the anti-correlation between the positive and negative subsectors.

What matrix structure results in the
subsector structure?
Let us consider a
$4\times4$ cross-correlation matrix,
\begin{equation}
C_{4\times4}=\left(
\begin{array}{cccc}
1 & 0.55 & 0.15 & 0.11\\
0.55 & 1 & 0.39 & 0.34\\
0.15 & 0.39 & 1 & 0.95\\
0.11 & 0.34 & 0.95 & 1
\end{array}\right),
\end{equation}
which is taken from the $\lambda_6$ sector of the GMI. The $1$-th and $2$-th indices represent
the 3-month and 6-month bond rates respectively, identified as the positive subsector.
The $3$-th and $4$-th indices are the 10-year and 20-year bond rates, identified as the negative subsector.
Obviously, the matrix elements $C_{ij}$ within the same subsectors, i.e., in the diagonal blocks, are
larger than the ones between the positive and negative subsectors, i.e., in the off-diagonal blocks.

To verify the anti-correlation more intuitively, therefore, we may calculate the average
$\overline C_{ij}$ within the positive or negative subsector,
and between the positive and negative subsectors.
The results for the NYSE are shown in figure~\ref{fig.2}, and those for the SSE are similar.
The average $\overline C_{ij}$ within the positive or negative subsector is obviously much
large than that between the positive and negative subsectors,
especially for small $\alpha$, i.e., large eigenvalues.
This strongly suggests that there
indeed exists an anti-correlation between the positive and negative subsectors.
However, we should keep in mind that the
anti-correlation in a particular eigenmode is only a part of the
cross-correlation between two stocks, as shown in
Eq~(\ref{eq:7}). How to make use of this anti-correlation theoretically and practically
remains challenging.

\section{Conclusion}

With the RMT method, we have investigated the spatial
structure of the SSE (China), NYSE (US) and GMI. Taking into account
the signs of the components in the eigenvectors of the
cross-correlation matrix, a sector may split into two subsectors, which
are anti-correlated each other in the corresponding eigenmode. The
results are shown in table~\ref{table-sector-SSE},
~\ref{table-sector-NYSE} and ~\ref{table-sector-GMI}. The NYSE is
dominated by the standard business sectors and subsectors, and the
GMI is controlled by the area sectors and subsectors, but without
the market mode. In contrast to it, the SSE exhibits unusual
sectors and subsectors.

\begin{figure}[t]
\includegraphics[scale=0.3]{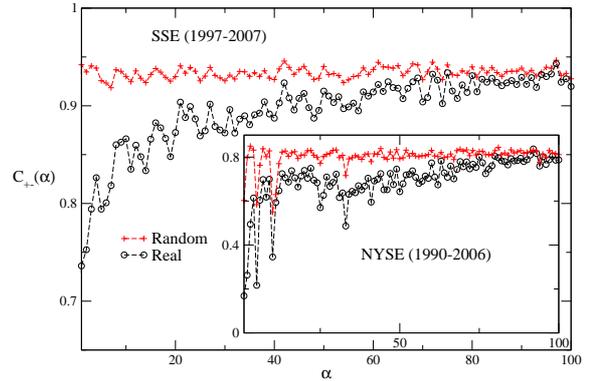}
 \caption{$C_{+-}(\alpha)$ for the SSE and NYSE are compared with that between two random combinations of stocks.}\label{fig.1}
\end{figure}

\begin{figure}[t]
\includegraphics[scale=0.3]{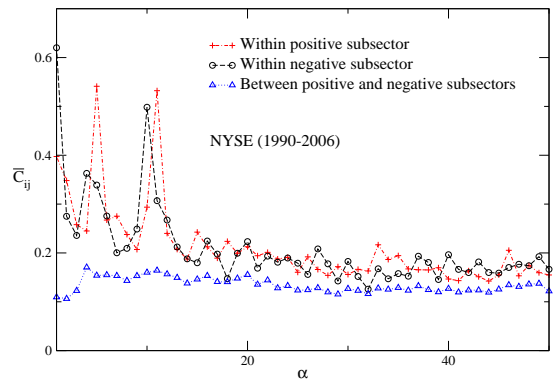}
\caption{The average cross-correlation
$\overline C_{ij}$ for the NYSE. } \label{fig.2}
\end{figure}

The subsector structure is strong in the
SSE, while somewhat weaker in the NYSE and GMI. The anti-correlation
between the positive and negative subsectors in an
eigenmode can be measured by $C_{+-}(\alpha)$ in Eq.~(\ref{eq:8}) and the average
$\overline C_{ij}$ within the positive or negative subsector,
and between the positive and negative subsectors, as shown in figures~\ref{fig.1} and \ref{fig.2}.

\acknowledgments This work was supported in part by NNSF of China
under Grant Nos. 10875102 and 11075137, and Zhejiang Provincial
Natural Science Foundation of China under Grant No. Z6090130.

\begin{table}
\caption{$\lambda_{min\left(max\right)}^{ran}$ denote the lower (upper) bound
of the eigenvalues of the Wishart matrix, while
$\lambda_{min}^{real}$, $\lambda_{0}$, $\lambda_{1}$ and
$\lambda_{2}$ represents the lower bound of the eigenvalues and the
three largest eigenvalues of the real systems respectively.}
\label{table-basic-stat}
 \centering \footnotesize
 \begin{tabular}{rccccccc}

 & $\lambda_{min}^{ran}$ & $\lambda_{max}^{ran}$ & $\lambda_{min}^{real}$ & $\lambda_{0}$& $\lambda_{1}$& $\lambda_{2}$ &  $\overline C_{ij} $ \tabularnewline
\hline SSE  & $0.47$ & $1.73$ & $0.18$ & $97.3$ & $4.17$ & $3.35$ &
$0.37$\tabularnewline NYSE  & $0.54$ & $1.55$ & $0.20$ & $45.6$ &
$8.71$  & $6.24$  & $0.16$\tabularnewline GMI & $0.72$ & $1.33$ &
$0.00$ & $21.5$ & $6.65$  & $5.40$ & $0.26$\tabularnewline \hline
\end{tabular}

\end{table}

\begin{largetable}
\caption{The subsectors in the SSE. The fraction is the
number of well identified stocks over the total number of stocks in the subsector.
Null: no obvious category; ST: specially treated; Trad: traditional
industry; Tech: high technology; SHRE: Shanghai real estate; Weak:
weakly cyclical industry; Stro: strongly cyclical industry; Fin:
finance; IG: industrial goods; Util: utility; Basic: basic
materials; Heal: health care; CG: daily consumer goods; Serv:
services.} \label{table-sector-SSE} \centering \footnotesize
\begin{tabular}{rcccccccccccc} \hline
 & \multicolumn{2}{c}{$\lambda_{1}$} & \multicolumn{2}{c}{$\lambda_{2}$} & \multicolumn{2}{c}{$\lambda_{3}$} & \multicolumn{2}{c}{$\lambda_{4}$}
&\multicolumn{2}{c}{$\lambda_{5}$} &
\multicolumn{2}{c}{$\lambda_{6}$}  \tabularnewline
\hline Sign & $+$
& $-$ & $+$ & $-$ & $+$ & $-$ & $+$ & $-$ & $+$ & $-$ & $+$ & $-$
\tabularnewline Sector & Null & ST & Trad & Tech & ST & SHRE & Weak
& Stro & Fin & Null & IG & Null \tabularnewline
$u_{c}^{\pm}=\pm0.08$ & $26$ & $ 31/35$ & $ 22/23$ & $ 23/25$ & $
24/27$ & $ 27/27$ & $23/26$ & $24/26$ & $14/18$ & $25$ & $15/17$ &
$25$ \tabularnewline $u_{c}^{\pm}=\pm0.10$ & $7$ & $ 20/23$ & $
16/17$ & $ 12/13$ & $ 11/12$ & $ 20/20$ & $ 13/15$ & $ 15/16$ & $
10/14$ & $17$ & $ 8/9$ & $18$ \tabularnewline \hline
\end{tabular}
\end{largetable}

\begin{largetable}
\caption{The subsectors in the NYSE. The abbreviations
can be seen in the caption of table \ref{table-sector-SSE}.}
\label{table-sector-NYSE} \centering \footnotesize

\begin{tabular}{rcccccccccccc}
\hline $\lambda_{i}$ & \multicolumn{2}{c}{$\lambda_{1}$} &
\multicolumn{2}{c}{$\lambda_{2}$} &
\multicolumn{2}{c}{$\lambda_{3}$} &
\multicolumn{2}{c}{$\lambda_{4}$} &
\multicolumn{2}{c}{$\lambda_{5}$} &
\multicolumn{2}{c}{$\lambda_{6}$} \tabularnewline \hline Sign & $+$
& $-$ & $+$ & $-$ & $+$ & $-$ & $+$ & $-$ & $+$ & $-$ & $+$ & $-$
\tabularnewline Sector & Util & Tech & CG & Basi & Tech & CG & Null
& Null & Basi & Null & Null & Fin \tabularnewline
$u_{c}^{\pm}=\pm0.08$ & $ 26/26$ & $ 3/4$ & $ 9/16$ & $ 23/26$ & $
15/26$ & $ 14/32$ & $26$ & $25$ &  $ 6/6$ & $9$ & $14$ & $16/20$
\tabularnewline $u_{c}^{\pm}=\pm0.10$ & $ 25/25$ & $ 0/0$ & $ 0/0$ &
$ 19/21$ & $ 6/13$ & $ 13/19$ & $18$ & $12$ & $ 0/0$ & $8$ & $8$ & $
16/18$ \tabularnewline $u_{c}^{\pm}=\pm0.12$ & $ 21/21$ & $ 0/0$ & $
0/0$ & $ 19/21$ & $ 5/7$ & $ 5/6$ & $7$ & $6$ &  $ 0/0$ & $4$ & $2$
& $ 8/11 $ \tabularnewline \hline

$\lambda_{i}$ & \multicolumn{2}{c}{$\lambda_{7}$} &
\multicolumn{2}{c}{$\lambda_{8}$} &
\multicolumn{2}{c}{$\lambda_{9}$} &
\multicolumn{2}{c}{$\lambda_{10}$}&
\multicolumn{2}{c}{$\lambda_{11}$}& \tabularnewline \hline Sign  &
$+$ & $-$ & $+$ & $-$ & $+$ & $-$ & $+$ & $-$ & $+$ & $-$ &&
\tabularnewline Sector & Tech & Serv & Serv & Heal & Fin & Serv &
Null & CG & Null & Serv&& \tabularnewline $u_{c}^{\pm}=\pm0.08$ & $
11/24$ & $ 12/29$ & $ 9/19$ & $ 9/16$ & $ 11/25$ & $ 13/24$ & $12$ &
$ 2/4$ & $7$ & $ 9/12$&& \tabularnewline $u_{c}^{\pm}=\pm0.10$ & $
7/13$ & $ 10/18$ & $ 8/11$ & $ 8/13$ & $ 6/11$ & $ 11/19$ & $6$ & $
2/3$ & $6$ & $ 9/10$&& \tabularnewline $u_{c}^{\pm}=\pm0.12$ & $
4/7$ & $ 8/11$ & $ 5/5$ & $ 7/7$ & $ 4/6$ & $ 10/10$ & $2$ & $ 2/3$
& $1$ & $ 7/8$&& \tabularnewline \hline
\end{tabular}

\end{largetable}

\begin{table}[h]
\caption{The subsector structure in the GMI . The
thresholds are $u^\pm_{c}=\pm0.15$. The bold Italic items are those
not belonging to the areas. NorA refers to the North America, and b3m and b1y
are the $3$ month and $1$ year bonds.}

\label{table-sector-GMI} \centering \footnotesize
\begin{tabular}{p{0.5cm}p{0.6cm}p{5.2cm}p{0.8cm}}
\hline
 & { Sign} &  & Area\tabularnewline
{ $\lambda_{0}$} & & {VLIC XMI DJA DJI DJT DJX IIX IXIC MID NDX NWX
}  & US\tabularnewline && { OEX PSE RUA RUI RUT SML SPC
}\textbf{\textsl{{ GDAXI GSPTSE }}}&\tabularnewline { $\lambda_{1}$}
& & { AORD HSI HSNC HSNF HSNP HSNU JKSE KS11 N225 NZ50 PSI STI}
\textbf{\textit{ ATX}}& Asia  \tabularnewline {$\lambda_{2}$} &  &
b6m b1y b2y b3y b5y b7y b20y b20y & Bond\tabularnewline
{$\lambda_{3}$} & + & AEX BFX FCHI FTSE GDAXI MIBTEL SSMI &
EU\tabularnewline
 & - & SHA SZA SHB SZB & China\tabularnewline
{$\lambda_{4}$} & + & HSI HSNC HSNP & HK\tabularnewline
 & - & AEX BFX FCHI FTSE GDAXI MIBTEL SSMI & EU\tabularnewline
 &&\textbf{\textit{SHA SZA SHB SZB}} & \tabularnewline
{$\lambda_{5}$} & + & IIX IXIC NDX NWX PSE SOXX & US\tabularnewline
 & - & XMI DJA DJI DJU DJX \textbf{\textit{b3m}} & US\tabularnewline
{$\lambda_{6}$} & + & b3m b6m b1y & Bond\tabularnewline
 & - & b7y b10y b20y \textbf{\textit{HSNU }} & Bond\tabularnewline
{$\lambda_{7}$} & + & AORD JKSE KS11 N225 NZ50 PSI TWII  &
Asia\tabularnewline
 & - & HSI HSNC HSNF HSNP HSNU \textbf{\textit{b3m }} & HK\tabularnewline
{$\lambda_{8}$} & + & BVSP IPSA MERV MXX \textbf{\textit{XAX
GSPTSE}} & NorA\tabularnewline \hline
\end{tabular}

\end{table}

\end{document}